\documentclass[prb,preprint]{revtex4-1} 

\usepackage[english]{babel}
\usepackage[utf8]{inputenc}
\usepackage{amsmath}
\usepackage{graphicx}
\usepackage{amssymb}
\usepackage[colorinlistoftodos]{todonotes}
\usepackage{hyperref}

\begin{document}

\title{Effects of Reducing Scaffolding in an Undergraduate Electronics Lab}

\author{Evan Halstead}
\email{ehalstea@skidmore.edu}
\affiliation{Physics Department, Skidmore College, Saratoga Springs, NY 12866}

\date{\today}

\begin{abstract}
Design and scientific investigation are recognized as key components of undergraduate physics laboratory curricula.  In light of this, many successful lab programs have been developed to train students to develop these abilities, and students in these programs have been shown to exhibit a higher transfer rate of scientific abilities to new situations.  In this paper, I use data from an electronics class for physics majors to investigate how giving students the opportunity to design circuits --- by removing steps from traditional cookbook lab guides --- affects the students' ability to determine the function of circuits they haven't seen before.  I compared post-lab quiz results from students who were given explicit procedures to those of students who were given incomplete procedures, and I found no statistically significant difference in the results of the two groups.  I explore possible explanations for the null effect and recommend future research directions.    
\end{abstract}

\maketitle

\section{Introduction}
\label{sec:Introduction}

There is currently widespread acceptance among the Physics Education Research community as well as organizations concerned with preparing students for future jobs in STEM fields that undergraduate laboratory curricula should have a strong focus on construction of knowledge and design of experiments to test models.\cite{AAPTlabguide, olson2012engage, gott1999practising, duggan2002sort, george1996shaping, czujko2000physics, scans1991work, bransford1999people, national1996national}  Many successful science curricula --- such as the Investigative Science Learning Environment (ISLE), Learning by Design (LBD), and Scientific Community Labs (SCL) --- have been developed with these as primary goals.\cite{etkina2007investigative, kolodner2002facilitating, redish2009reinventing}  Etkina et al showed that ISLE students who designed their own experiments not only performed as well as other non-design ISLE students on traditional exams, but they also significantly outperformed the non-design group on experimental tasks that required application of scientific abilities to a new situation.\cite{etkina2010design,etkina2007spending, karelina2007design}  

A key component of curricula like ISLE and LBD is the implementation of an inquiry learning cycle and having the opportunity to practice that learning cycle many times.  That is, students are provided with a framework that reflects scientific habits of mind.\cite{etkina2007investigative, etkina2006using, kolodner2002facilitating, ryan2001design}  Many examples of such frameworks can be found in the literature.\cite{zwickl2014incorporating, holmes2015teaching, dounas2015role}  While these learning cycles tend to focus on designing experiments to answer a particular question (i.e., What is the relationship between the period and length of a pendulum?), another class of design --- which is perhaps more relevant to this study --- is what could be called \textit{engineering design}.  That is, students design a device or system capable of achieving a certain goal (i.e., a circuit capable of turning on an LED two seconds after a switch is thrown).  The Engineering is Elementary (EiE) curriculum has developed their own learning cycle for this kind of design.\cite{cunningham2007engineering}  What all of the learning cycles have in common is that students develop a plan, execute the plan, compare results against expectations/desired results, then go back and refine the plan.  The process continues until the students have answered the experimental question with sufficient precision or achieved the engineering design goal.  Lawson argues that the average student (at least by college age) has the capacity to understand such a process,\cite{lawson2003nature} and Hammer and Elby go further and claim that students already use such a framework (which they call an \textit{epistemological resource}) in other contexts in their lives.  For example, a familiar design problem is one of arranging furniture in a room; it is not uncommon for people to imagine several possible arrangements, try each one, then finally settle on one choice.\cite{hammer2003tapping}  The ISLE program uses a version of this cycle, and after several iterations students begin to internalize the process.\cite{etkina2008long}  

Unfortunately, making significant changes to classroom practices is difficult.  Henderson and Dancy suggest that even when teachers have views of teaching that are compatible with those recommended in the literature, their actual teaching practices may not match their own views because of situational factors.\cite{henderson2007barriers}  For example, new teachers and/or teachers with other significant time commitments may not be able to undertake a lab curriculum overhaul, especially if there aren't any currently available resources that are appropriate for his or her class.  It is worth asking, then, whether there is anything to be gained from modifying lab guides to encourage students to employ the reasoning processes emphasized by curricula such as those mentioned above but without devoting the time and resources to explicitly training students to do so. While there are many possible benefits, this paper will focus on one in particular: does reducing scaffolding produce a better ability to explain and model novel circuits?   

This paper is organized as follows: In Sec. \ref{sec:Methods}, I discuss the experimental methods.  In Sec. \ref{sec:Results}, I discuss the results and some of the challenges I encountered.  Finally, Sec. \ref{sec:Conclusions} offers concluding remarks.  

\section{Methods}
\label{sec:Methods}

In the Spring of 2015, I taught an undergraduate electronics course (N=20) at a small liberal arts college.  The majority of the students were sophomore and junior physics majors and minors.  Electronics is a subject with a wide range of audiences, including hobbyists, technicians, and engineers, who are often less interested than physicists in \textit{why} things work in a particular way.  Therefore, many of the readily available resources for circuit-building lab activities, such as \url{www.allaboutcircuits.com}, focus on following a rote procedure.\cite{allaboutcircuitsexperiments}  The lab guides I used, then, were a combination of labs of my own design as well as labs taken and modified from \textit{Electronics with Discrete Components} by Enrique J. Galvez.\cite{Galvezbook}  These labs were designed to test the difference between cookbook lab guides and lab guides where students are given the opportunity to design their own circuits.    

There were five total lab activities that were chosen for this investigation.  In all cases, students were asked to design and build circuits to achieve a specified task (i.e., turn on a light with a two-second delay, display the binary sum of two one-digit numbers, build a human touch sensor).  The tasks required information that had been discussed previously in class, but their notes did not contain direct solutions to the tasks.  In each activity, students in the class were separated into six groups consisting of three or four students.  Each group sat at its own table that was physically separated enough from the other tables that there was minimal cross-talk between groups.  I then randomly split the six groups into two experimental groups.\footnote{Changing the groups each time had the effect of increasing the effective sample size, but it also prevented me from looking at long-term effects.}  The Maximally Scaffolded (MS) group received lab guides that explicitly outlined each step of the construction and analysis of a circuit; these were meant to be like traditional cookbook lab guides.  The Reduced Scaffolding (RS) group was given a similar lab guide, but some of the steps were deleted.  Examples are shown in Appendix A.  It should be noted that on the surface these two example lab guides look very similar because the Reduced Scaffolding lab only removed a \textit{portion} of the scaffolding.  The first reason for this was that my assessment of the students' understanding of the circuit always focused on one particular portion of the circuit.  The second reason was logistical: since students from each of the two experimental groups frequently walked around the room to gather materials, there was a significant chance that they would see each other's lab guides.  This meant that the lab guides had to superficially look the same so that students would be unaware of the experimental conditions.  

In order to assess the students' understanding of the circuit that they built, I gave each student a post-lab quiz before they left that day.  The quizzes for all of the lab activities followed the same general template so that I could more reliably compare cross-quiz results and so that I could check if there was a difference in performance on different $\textit{kinds}$ of questions.  Each quiz consisted of three two-part ``check all that apply'' questions.  The first part of each question asked for an answer, and the second part of each question asked for an explanation of why the first part's answer was chosen.  This was done to differentiate correct guesses from correct reasoning.  The three quiz questions were structured as follows.

\begin{enumerate}
\item
Circuit Recall (CR): The first question in each post-lab quiz generally tested the student's ability to recall the details of the circuit.  In some lab activities, digital Integrated Chips (ICs) were used instead of analog circuit components, and a circuit diagram wasn't a feasible option for the CR question.  In those cases, the question instead tested for recall of a different visual element such as a truth table.
\item
Procedural Recall (PR): The second question in each post-lab quiz generally tested the student's ability to recall a particular procedure or operation that was carried out during the lab activity.  This was usually an equation or Boolean operation.
\item
Altered Procedure (AP): The third question in each post-lab quiz asked the student to determine the function of a circuit that was slightly different from the one they constructed in lab.  This usually involved switching the order of components or adding/removing components. 
\end{enumerate}
An example quiz is provided in Appendix B.  I included the CR and PR questions because I was interested in seeing whether the presence of a visual cue in the lab guide would help them later remember what they did.  Alternatively, perhaps having to design the circuit and look for the proper equations themselves would help the students remember better.  This effect is similar to what Schwartz et al saw in their research study on Tell \& Practice versus Inventing with Contrasting Cases.  Middle-school students learning about density were shown pictures of buses with different lengths that were filled with different numbers of clowns.  One group of students was told the formula for density and asked to practice using the buses and clowns.  Another group was asked to invent a ``crowdedness'' index.  On the following day, students were asked to redraw the activity from the day before; the researchers found that students who invented their own formulas had better recall of deep structural features of the activity than students who were told the formulas.\cite{schwartz2011practicing}  The AP question was included because students designing their own experiments spend more time sense-making.\cite{karelina2007and, lippmann2002analyzing}  I therefore hypothesized that the students in the Reduced Scaffolding group would perform better on the AP questions since they would be more likely to consider several designs for their circuit.  In the process of considering those designs, they may have come across the actual design presented in the question or may have considered enough designs that they would be better equipped to speculate about how changes would affect function.   

\section{Results and Discussion}
\label{sec:Results}
In order to separate out the effects of the MS and RS treatments, I performed a mixed-effects logistic regression\cite{lme4package} of the form
\begin{equation}
\text{Score}_{ijk}=\beta_{1k}\times\text{Quiz}_k+\beta_{2j}\times\text{Treatment}_j+\epsilon_i,
\end{equation}
where $\text{Score}_{ijk}$ is the quiz score of student $i$ in treatment group $j$ on quiz $k$, $\text{Quiz}_k$ is a fixed-effect variable that accounts for the difficulty of the quiz, $\text{Treatment}_j$ is a fixed-effect variable that is either MS or RS, and $\epsilon_i$ is a random intercept that accounts for the variability of each student.  This type of analysis can be used in experiments with a crossover design,\cite{ives2014measuring} and it has the benefit of not treating the quizzes as identical.  For any given quiz and student, a difference in the components of the $\beta_2$ vector would mean a predicted difference in the score depending on the treatment group.  I found, however, that there was no statistically significant difference between the MS and RS groups.  This was true both when I used the total quiz score as well as when I looked at the scores of individual quiz questions.

\begin{figure}[h!]
\includegraphics[width=\textwidth]{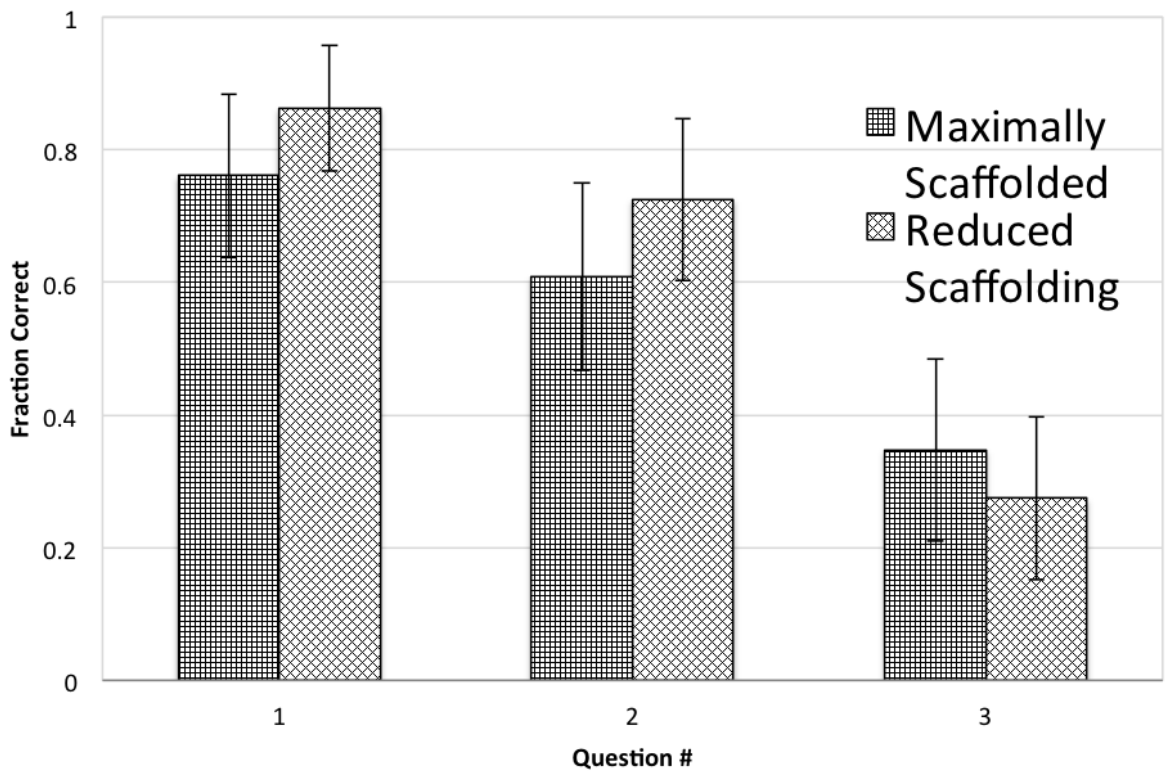}
\caption{\label{fig:Results_AllQuestions}These are the aggregated results for each of the three questions from all of the post-lab quizzes.  The question numbers correspond to the CR, PR, and AP questions, respectively.  Error bars are $2\sigma$.}
\end{figure}

While there was no statistically significant difference between the MS and RS groups, one interesting result is that there was a significant drop-off in performance for both groups on the AP question when the results from all five quizzes were averaged together, as shown in Fig.\ref{fig:Results_AllQuestions}. 

These results, of course, do not prove that there was no difference in performance between the MS and RS groups but rather that the experiment failed to prove that there was a difference.  Perhaps with a larger sample size there would have been an observable difference.  Still, it is worth speculating as to why there wasn't more of a difference, especially for the AP questions.  Note that benefits of RS-type lab tasks were measured through performance on post-lab quizzes; it is possible that despite similar performance on test questions, RS students were in fact employing a more scientifically minded methodology than MS students. Etkina et al, when testing the effects of design versus non-design labs in ISLE, found that students who performed design labs scored similarly on exams to those who did not design their own labs, but did better on assessments of scientific abilities when working on novel experimental tasks.\cite{etkina2010design, karelina2007design} This may be a result of students’ activation of certain reasoning abilities being context-dependent. \cite{hammer2003tapping} For example, in some situations students may see knowledge as something that comes from the teacher or textbook while in other situations they may instinctively try to figure things out by using the knowledge that they already have.  From this view, then, it may be that even the RS students were employing ``knowledge as coming from a teacher or textbook'' when working on the AP questions.  This seems plausible to me, as the most common incorrect answer to \#3 in the example quiz in Appendix B was ``the order of circuit elements that are in series with one another does not matter'', which is a result of a commonly used heuristic which does not apply in this question.  In fact, answers based on heuristics were very common in all of the AP questions, which indicates that students may have been answering based off of rules of thumb that they remembered rather than trying to figure out the answers.  Of course, it is also possible that the single exposure to employment of design methodologies in physics contexts is insufficient to produce significant changes in RS students’ approaches to AP-style questions. Further testing would be required to determine whether RS and MS students did in fact employ different ways of reasoning when answering post-lab questions.         

Finally, I wish to comment on anecdotal observations and logistical challenges.  One difference I did observe was that on average students in the MS group finished the lab activities much more quickly than students in the RS group, which is consistent with what Karelina and Etkina observed in ISLE classes.\cite{karelina2007acting}  Students in the RS group often worked up until or even past the end of the scheduled class period.  This may have affected the quiz results because those students had more incentive to rush through the quiz.\footnote{My grading system allowed students to correct quizzes and homeworks in one five-minute screencast each week, so the only downside to rushing through a quiz was that it would take away from the five minutes that they could dedicate to correcting homework instead.}  It would be interesting to see if the quiz results would be any different if the time between the lab activity and the quiz had been longer, but I do not have that data.  

One challenge, as mentioned earlier, was that students in both experimental groups used the same room at the same time.  This presented the possibility of collaboration between groups and even recognition that other groups in the room had been given a different experimental procedure.  Anecdotally, I did not observe inter-group collaboration.  When students did encounter difficulties, they either proceeded by trial-and-error or they asked me for help (in which case I helped only by asking leading questions.)  A debriefing poll after the course ended also revealed that none of the students had caught on to the true nature of the experiment.    

\section{Conclusions}
\label{sec:Conclusions}

It is currently recognized that undergraduate laboratory curricula should focus not just on measurement techniques but also experimental design.  This represents a departure from the style of traditional lab activities in which students are provided with a list of steps to carry out in order to verify a certain result.  To address the experimental design component, several education research groups have recently worked to develop laboratory curricula with learning cycles that train students in how to design their own experiments, and these curricula have been very successful according to the research literature.  Implementing these techniques, however, requires some lab restructuring and training on the part of the instructor.  This could act as a barrier to entry for some instructors.  I was therefore interested in determining whether simple modifications to lab guides could result in benefits to student learning.

Students in an Electronics course for physics majors at a small liberal arts college were randomly placed into one of two experimental groups for each of five lab activities.  The Maximally Scaffolded group was given an explicit set of steps to construct a circuit to achieve a certain task while some of the steps were removed from the lab guides for students in the Reduced Scaffolding group.  Post-lab quizzes assessed recall of the circuit, recall of procedural operations, and understanding of the function of a slightly altered circuit.  Results showed no statistically significant difference between the two groups on the post-lab quiz.  
 
The lack of a statistically significant difference on post-lab quiz scores does not mean that there is no difference. Further experiments would be needed to determine whether RS-type lab guides are sufficient to increase student’s likelihood of using a scientifically minded reasoning methodology rather than relying simply on recall or whether explicit instruction on such methods is required.  One could also record student interactions to look for differences in the way that the students approach solving the design problems. This would be a more direct assessment of the design process itself.     

The results of this experiment show that simply deleting steps from prototypical ``cookbook'' lab guides may not be sufficient to increase students’ takeaway from lab activities. 

\appendix

\section{Sample Lab Guides (MS and RS)}
\begin{center}
\textbf{Delayed Switch}
\end{center}
\textbf{Required equipment:} power supply

\noindent \textbf{Required components:} SPDT switch, CD4066 digital switch, LED, resistors, capacitors

Often we want to take a group photo, but we want everybody in the picture.  A delayed shutter is the solution.  Here we use a circuit similar to the last lab, but we need to delay the closing of the switch by 2 seconds (a fast run to join the group).  Check the schematic of the circuit in the figure below.  You must design a capacitor-charging circuit that triggers the switch ($V_C>2.5\text{ V}$ at pin 13) 2 seconds after the manual switch is closed.  

\begin{enumerate}
\item
Starting with the charging equation for a capacitor in an RC circuit, $V=V_0(1-e^{-t/RC})$, solve the expression for the resistance $R$.
\item
Find a capacitor and take note of its capacitance.  Based on the description above, determine the resistance by substituting the appropriate values into the expression you determined in step 1.  Find the resistor that most closely matches your calculated number.
\item
Assemble the circuit shown below and test it.
\end{enumerate}

\begin{figure}[h!]
\includegraphics[width=\textwidth]{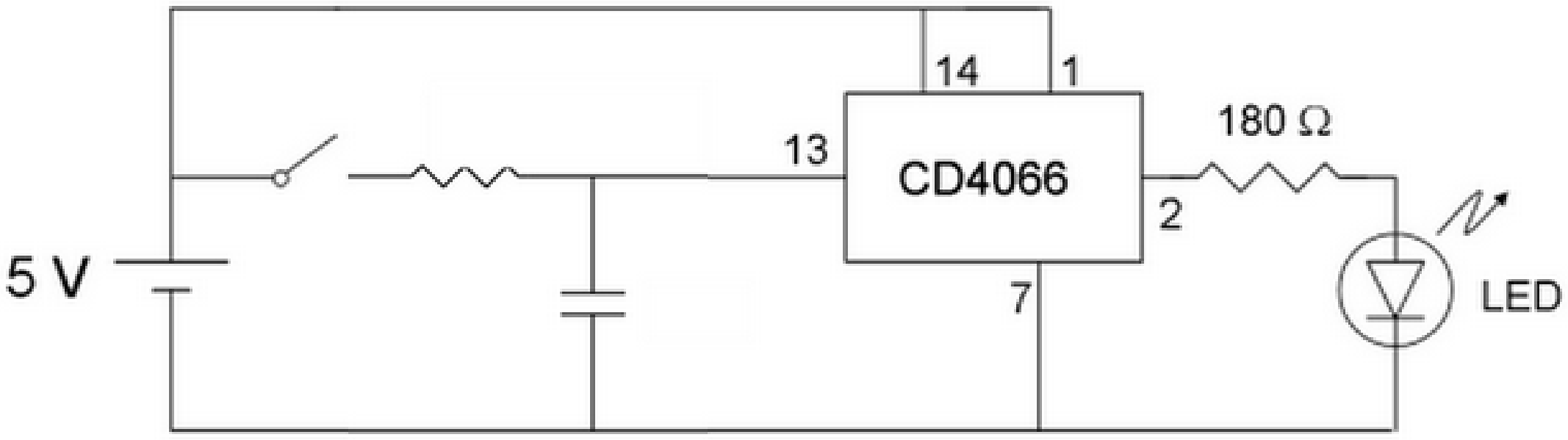}
\end{figure}

\vspace{3cm}
\begin{center}
\textbf{Delayed Switch}
\end{center}

\textbf{Required equipment:} power supply

\noindent \textbf{Required components:} SPDT switch, CD4066 digital switch, LED, resistors, capacitors

Often we want to take a group photo, but we want everybody in the picture.  A delayed shutter is the solution.  Here we use a circuit similar to the last lab, but we need to delay the closing of the switch by 2 seconds (a fast run to join the group).  Check the schematic of the circuit in the figure below.  You must design a capacitor-charging circuit that triggers the switch ($V_C>2.5\text{ V}$ at pin 13) 2 seconds after the manual switch is closed.  When deciding values for components, first see the capacitor values that are available, as we have a larger variety of resistors than capacitors.

\begin{figure}[h!]
\includegraphics[width=\textwidth]{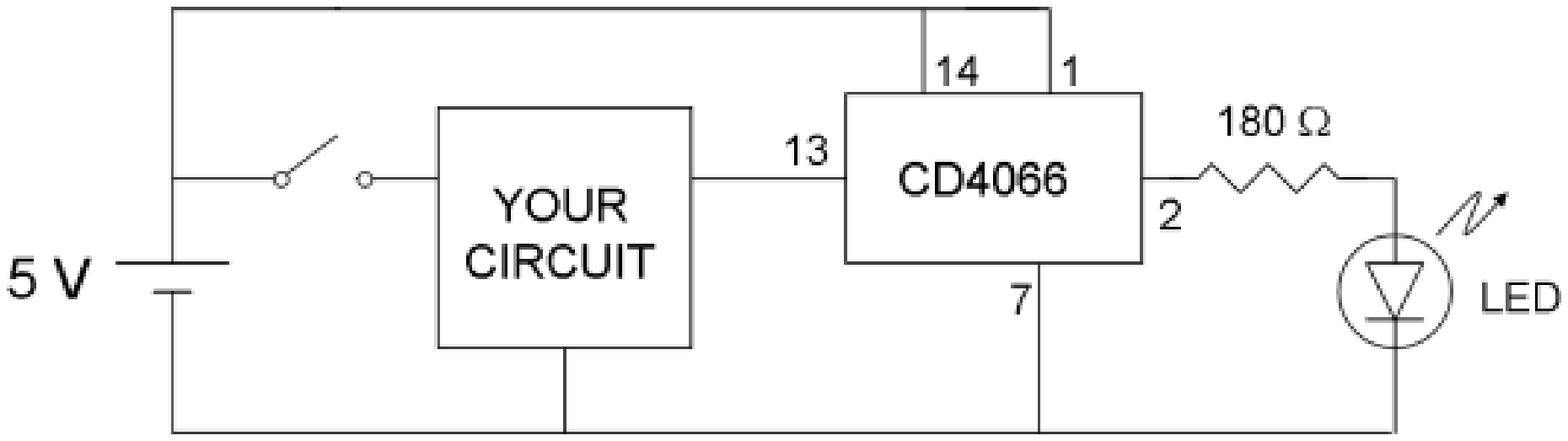}
\end{figure}

\section{Post-Lab Quiz Example}
\begin{enumerate}

\item
The figure below shows the circuit that you used in today's circuit-building activity.  

\begin{figure}[h!]
\includegraphics[width=8cm]{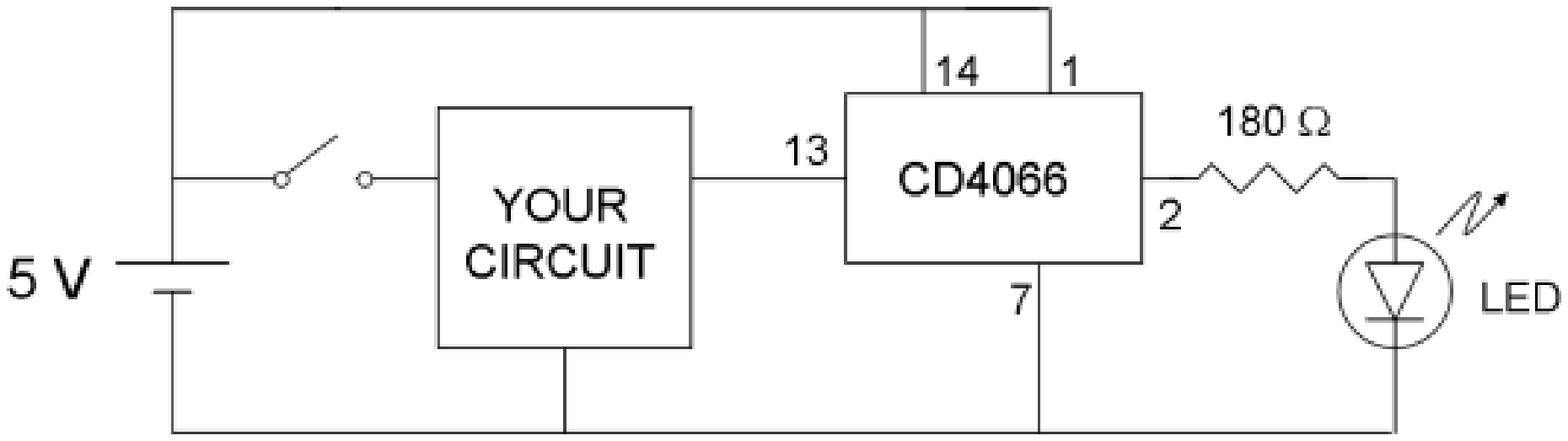}
\end{figure}

Circle the schematic that appropriately indicates the arrangement of capacitor and resistor that you used for ``Your Circuit''.

\begin{figure}[h!]
\includegraphics[width=6cm]{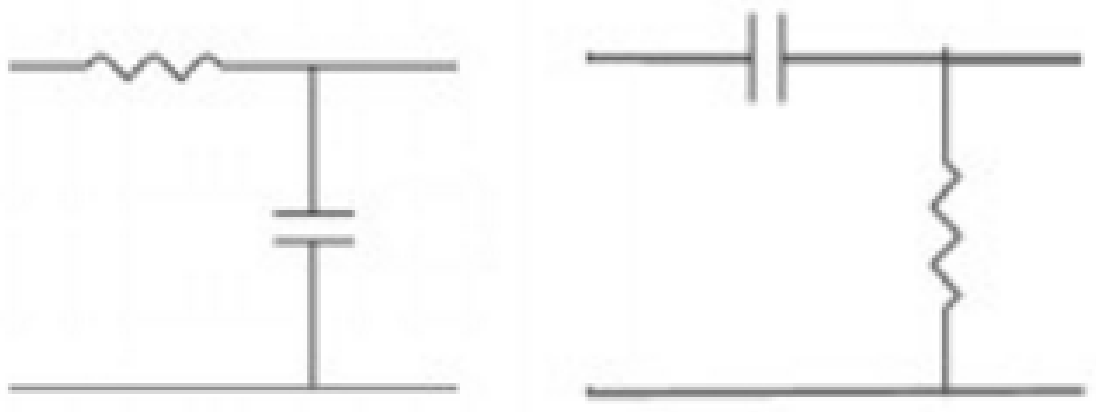}
\end{figure}

This particular arrangement was used because:

$\Box$ the capacitor discharges after the switch is closed and eventually reaches the 2.5 V at pin 13 that is necessary to activate the CD4066.

$\Box$ the capacitor charges after the switch is closed and eventually reaches the 2.5 V at pin 13 that is necessary to activate the CD4066.

$\Box$ the capacitor discharges after the switch is closed and eventually reaches the 0 V at pin 13 that is necessary to activate the CD4066.

$\Box$ the capacitor charges after the switch is closed and eventually reaches the 5V at pin 13 that is necessary to activate the CD4066.

\item
What equation did you use to eventually determine the values of resistance and capacitance that you needed to create a 2s time delay?

$\Box V=V_0e^{t/RC}$

$\Box V=V_0e^{-t/RC}$

$\Box V=V_0(1-e^{-t/RC})$

$\Box V=V_0(1-e^{t/RC})$

You used this equation because:

$\Box$ it is the equation for the voltage across a charging capacitor.

$\Box$ it is the equation for the voltage across a discharging capacitor.

$\Box$ it is the equation for the charge on a charging capacitor.

$\Box$ it is the equation for the charge on a discharging capacitor.

\item
If you were to reverse the order of the capacitor and resistor before closing the switch:

$\Box$ the circuit would still work the same.

$\Box$ the circuit would not work as desired.

$\Box$ the LED would take a longer amount of time to turn on.

$\Box$ the LED would take a shorter amount of time to turn on.

$\Box$ the LED would turn on 2 seconds before you closed the switch rather than 2 seconds after.

This is because:

$\Box$ the voltage has to get through the capacitor first.

$\Box$ the voltage has to get through the resistor first.

$\Box$ the order of circuit elements that are in series with one another does not matter.

$\Box$ the voltage at pin 13 would start high and decrease over time.

$\Box$ the voltage at pin 13 would start low and increase over time.

$\Box$ switching the order of the capacitor and resistor would make time go backwards in the circuit.

\end{enumerate}

\begin{acknowledgments}
The author would like to thank the reviewers for their many helpful comments as well as Michael Lopez from the Skidmore College Mathematics Department for consultation on the statistical analysis.
\end{acknowledgments}

\end{document}